\documentstyle[prd,aps]{revtex}
\def\g{\gamma} 
\def\a{\alpha} 
\def\b{\beta} 
\def\l{\lambda} 
\def\m{\mu} 
\def\p{\phi}
\def\vp{\varphi}
\def\r{\rho} 
\def\t{\tau} 
\def\io{\iota}
\def\o{\omega} \def\O{\Omega}
\def\we{\wedge} 
\def\pr{\prime}
\def\be{\begin{equation}}
\def\ee{\end{equation}}

\begin{document}
\draft

\title{ ``Self Dual" Solutions of Topologically 
Massive Gravity Coupled with the Maxwell-Chern-Simons Theory}

\bigskip
\author{T. Dereli and  \"{O}. Sar{\i}o\u{g}lu} 
\address{Department of Physics,
Middle East Technical University,
06531 Ankara, TURKEY}

\maketitle

\bigskip
\begin{abstract}
We give a general class of exact solutions to the $(1+2)$-dimensional
topologically massive gravity model coupled with Maxwell-Chern-Simons
theory where a ``self duality" condition is imposed on the Maxwell 
field. 
\end{abstract}

\vskip 1cm

It is well known that general relativity in $(1+2)$ dimensions has
no degrees of freedom and the gravitational field is determined
solely by the matter sources (see \cite{carlip} and the references 
therein). However a dynamical model is provided by the topologically
massive gravity (TMG) theory which is obtained by the addition of the 
gravitational Chern-Simons term to the usual Einstein-Hilbert piece
in the action \cite{djt}. Recently, a general class of exact black
hole solutions to TMG with a negative cosmological constant has been
obtained \cite{DS} and it has also been shown that these solutions 
are supersymmetric and asymptotically approach the extremal BTZ black hole 
solution \cite{BTZ}. Here, we present a new class of solutions 
to TMG coupled with Maxwell-Chern-Simons theory where the Maxwell 
field has been constrained to obey the ``self duality" condition.

We begin with the action \( I[e,\o,A] = \int_{M} L \) where the 
Lagrangian 3-form is given by
\be
L = \frac{1}{\m} (\o^a_{\;b} \we d \o^b_{\;a} + \frac{2}{3} 
\o^a_{\;b} \we \o^b_{\;c} \we \o^c_{\;a}) + \frac{1}{2} {\cal R} * 1
- \l * 1 - \frac{1}{2} F \we * F - \frac{1}{2} m \, A \we F \;\; . 
\label{Lag}
\ee
Apart from the usual Einstein-Hilbert term and the negative
cosmological constant $\l=-1/l^2 < 0$, the gravitational Chern-Simons
term with the coupling constant $\m$, which has dimensions of mass,
is written in terms of the connection 1-forms $\o^a_{\;b}$; there 
is also the standard Maxwell Lagrangian given in terms of the Maxwell
field $F \equiv dA$ along with the vector Chern-Simons term with 
the coupling constant $m$. The variation of $I$ with respect to the 
orthonormal coframes $e^a$ and the electromagnetic potential $A$ 
yields
\begin{eqnarray}
\frac{1}{\m} C_a + G_a + \l * e_a & = & - \t_a[A] \;\; , 
\label{gravfe} \\
d * d A + m \, d A & = & 0 \;\; , \label{emfe}
\end{eqnarray}
respectively. Here 
\[ \t_a[A] = - \frac{1}{2} (\io_a d A \we * d A - 
               d A \we \io_a * d A) \]
is the electromagnetic field energy momentum 2-form along with the
standard Einstein 2-forms \( G_a \equiv G_{ab} * e^b = - \frac{1}{2}
R^{bc} * e_{abc} \) and the Cotton 2-forms \( C_a \equiv D Y_a =
d Y_a + \o_a^{\;b} \we Y_b \), where \( Y_a \equiv (Ric)_a - 
\frac{1}{4} {\cal R} e_a \) is defined in terms of the Ricci 1-forms
\( (Ric)_b \equiv \io_a R^a_{\;b} \) and the curvature scalar 
\( {\cal R} \equiv \io_a (Ric)^a \). The $R^a_{\;b}$, of course, are
the curvature 2-forms \( R^a_{\;b} = d \o^a_{\;b} + \o^a_{\;c} \we
\o^c_{\;b} \) of the connection 1-forms $\o^a_{\;b}$, which satisfy
the Cartan structure equations \( d e^a + \o^a_{\;b} \we e^b = 0 \),
so that there is no torsion present in the theory. The Hodge duality
operation is specified with the oriented volume element \( * 1 =
e^0 \we e^1 \we e^2 \) and $e_{abc}$ is a short hand notation for
$e_a \we e_b \we e_c$.

We choose the local coordinates $(t,\r,\p)$ and make a general ansatz
for the Maxwell field
\be
F = d A = E(\r) \, e^0 \we e^1 + B(\r) \, e^1 \we e^2  \;\; , \label{F=dA}
\ee
and for the orthonormal coframe 1-forms as
\be 
e^0 = f(\r) dt \;\;\;\; , \;\; e^1 = d \r \;\;\;\; , \;\;
e^2 = h(\r) (d \p + a(\r) dt) \;\; , \label{metes}
\ee
so that the metric takes the form 
\be 
ds^2 = - e^0 \otimes e^0 + e^1 \otimes e^1 + e^2 \otimes e^2 \;\; , \label{met}
\ee
which is most suitable for a study of rotationaly symmetric solutions.

This choice gives
\begin{eqnarray}
-Z + \frac{1}{l^2} + \frac{1}{\m} \left( X^{\pr} + 
\g X + \frac{1}{2} \b (Y-W) \right) 
& = & \frac{1}{2} (E^2 + B^2)  \;\; , \label{gravz} \\
-X + \frac{1}{\m} \left( \frac{1}{2} (Z-W-Y)^{\pr} + \a (Z-Y) 
+ \frac{3}{2} \b X \right) & = & - E B  \;\; , \label{gravx} \\
Y - \frac{1}{l^2} + \frac{1}{\m} \left( (\g-\a) X + \frac{1}{2} \b (W-Z) 
\right) & = & - \frac{1}{2} (E^2 - B^2)  \;\; , \label{gravy} \\
W - \frac{1}{l^2} + \frac{1}{\m} \left( X^{\pr} + \a X + \frac{1}{2} \b 
(Y+Z-2W) \right) & = & \frac{1}{2} (E^2 + B^2)  \;\; , \label{gravw} 
\end{eqnarray}
for the gravitational field equations (\ref{gravfe}),
whereas the modified Maxwell equations (\ref{emfe}) become:
\begin{eqnarray}
B^{\pr} + \a B - (\b+m) E & = & 0 \;\; , \label{maxb} \\
E^{\pr} + \g E - m B & = & 0 \;\; , \label{maxe} 
\end{eqnarray}
where we denote derivatives with respect to $\r$ by a prime.

Here we introduced the functions:
\be
W \equiv \a^{\pr} + \a^2 - \frac{3}{4} \b^2 \;\;\;\; , \;\;
X \equiv \frac{1}{2} \b^{\pr} + \g \b \;\;\;\; , \;\;
Y \equiv \a \g + \frac{1}{4} \b^2 \;\;\;\; , \;\;
Z \equiv \g^{\pr} + \g^2 + \frac{1}{4} \b^2 \;\;\;\; , \label{wxyz}
\ee
which actually describe the curvature 2-forms $R^a_{\;b}$, and
\be
\a \equiv \frac{f^\pr}{f} \;\;\;\; , \;\; \b \equiv \frac{a^\pr h}{f} 
\;\;\;\; , \;\; \g \equiv \frac{h^\pr}{h} \;\;\;\; , \label{albega}
\ee
that come from the connection 1-forms $\o^a_{\;b}$ (see \cite{DS} 
for details). 

Assuming the ``self duality" of the electromagnetic field as
\be
E = k B  \;\;\;\; {\rm with} \;\; k^2 = 1 \label{sedual}
\ee
and substituting this into (\ref{maxb}) and (\ref{maxe}), we find
$\g + k \b = \a$, and using this in (\ref{gravy}) gives $ (\g + 
\frac{k}{2} \b)^2 = \frac{1}{l^2}$. Hence one finds that $\a$ and 
$\g$ are determined by $\b$ alone as
\be
\a = \frac{k}{2} \b + \frac{1}{l} \;\;\; , \;\; 
\g = - \frac{k}{2} \b + \frac{1}{l}  \;\;.  
\label{algaitobe}
\ee
These ubiquitous conditions have first appeared in the study of the 
general self dual solutions of the Einstein-Maxwell-Chern-Simons theory
in (1+2) dimensions \cite{DerObu}. They are in fact the necessary and 
sufficient conditions that any solution to TMG of the form (\ref{metes}),
(\ref{met}) has to satisfy in order to be supersymmmetric as well \cite{DS}.

The above conditions (\ref{algaitobe}) simplify $W$, $X$, $Y$ and $Z$ 
which now take the form
\be
W = u + \frac{1}{l^2} \;\;\;\; , \;\; X = k u \;\;\;\; , \;\; 
Y = \frac{1}{l^2} \;\;\;\; , \;\; Z = \frac{1}{l^2} -u
\label{simwxyz}
\ee
where
\be
u \equiv - \frac{1}{2} \, \b^2 + \frac{k}{2} \, \b^{\pr} + k \, 
\frac{\b}{l} \;\; .
\label{defnu}
\ee
Finally, we find that equations (\ref{gravz})...(\ref{maxe}) are
satisfied simultaneously provided
\begin{eqnarray}
u^\pr - k \, \b \, u + (\frac{1}{l}+ \m k) u & = & y \;\; , \label{eqnu} \\
y^\pr - k \, \b \, y + 2 (\frac{1}{l} - m k) y & = & 0 \;\; , \label{eqny}
\end{eqnarray}
where we defined $y \equiv k \, \m \, B^2$.

By setting $y = k \frac{u}{z}$ in (\ref{eqny}), we find a linear
first order differential equation for $z$ as
\be
z^\pr + (\m k + 2 m k - 1/l) z = k \;\; . \label{eqnz}
\ee
Integrating this, one finds
\be
z = \frac{k}{\m k + 2 m k - 1/l} \; \left(
1 + y_0 e^{(1/l -\m k - 2 m k) \r} \right) 
\label{solnz}
\ee
for some integration constant $y_0$. Substituting this back into
(\ref{eqnu}), one finds that
\be
u^\pr - k \, \b \, u + \left( \frac{1}{l} + \m \, k -
\frac{\m k + 2 m k - 1/l}{1 + y_0 e^{(1/l -\m k - 2 m k) \r }} 
\right) u = 0
\ee
and taking $u = k \frac{\b}{v}$, this simplifies to a linear first
order differential equation for $v$ as
\be
v^\pr + \left( \frac{1}{l} - \m \, k + 
\frac{\m k + 2 m k - 1/l}{1 + y_0 e^{(1/l -\m k - 2 m k) \r} }
\right) v = 2 \;\; .
\ee
This can be integrated easily and one finds
\be
v = \frac{1}{e^{(1/l-k \m)\r} + \frac{1}{y_0} e^{2 k m \r}} \;
\left( v_0 + \frac{2}{1/l- \m k} e^{(1/l-\m k)\r} 
+ \frac{1}{k m y_0} e^{2 k m \r} \right) 
\ee
with a new integration constant $v_0$.

Going back to the definition of $u$, (\ref{defnu}), gives a 
differential equation for $\b$ and by setting 
$\o \equiv \frac{1}{\b}$, it becomes:
\be
\o^\pr + \left( 
\frac{2 e^{(1/l-\m k)\r} + \frac{2}{y_0} e^{2 k m \r}}
{v_0 + \frac{2}{1/l- \m k} e^{(1/l-\m k)\r}  
+ \frac{1}{k m y_0} e^{2 k m \r} } 
-\frac{2}{l} \right) \o + k = 0 \;\; .
\ee
This is of the form
\be
\o^\pr + \left( \frac{\vp^\pr}{\vp} - \frac{2}{l} \right) \; 
\o + k = 0 \;\;\; , \label{omeqn}
\ee
with
\be 
\vp = v_0 + \frac{2}{1/l- \m k} e^{(1/l-\m k)\r}  
+ \frac{1}{k m y_0} e^{2 k m \r} \;\;\; , \label{varphi}
\ee
and its solution is given by (see \cite{DerObu})
\be
\o = \frac{1}{\b} = \frac{k \O}{\vp e^{- 2 \r / l}} \;\;\; {\rm where} 
\;\;\; \O \equiv c_0 - \int^{\r} d \tilde{\r} \; \vp(\tilde{\r}) \; 
e^{-2 \tilde{\r} / l}
\ee
for some integration constant $c_0$. In this case, $\O$ is:
\be
\O = c_0 + \frac{v_0 l}{2} e^{-2 \r /l} + \frac{2}{1/l^2 - \m^2} 
e^{-(1/l + \m k)\r} - \frac{1}{2 k m y_0 (k m -1/l)} e^{2(k m -1/l)\r}
\;\;\; . \label{Omega}
\ee

Finally, integrating for the metric functions using (\ref{albega})
and (\ref{algaitobe}), we find
\begin{eqnarray}
f & = & f_0 \; e^{\r /l} \; \O^{-1/2} \;\;\; , \label{solf} \\
h & = & h_0 \; e^{\r /l} \; \O^{1/2} \;\;\; , \label{solh} \\ 
a & = & -a_0 + k \; \frac{f_0}{h_0} \; \O^{-1} \;\;\; , \label{sola}
\end{eqnarray}
for some integration constants $f_0$, $h_0$, $a_0$ whereas the magnetic
field becomes
\be
B^2 = \frac{\m k + 2 m k - 1/l}{k \m y_0} \; e^{2(k m-1/l)\r} 
\; \O^{-1} \; .
\ee

As a first check of this solution, we look at the 
\( k \m \rightarrow \infty\) limit. Then the contributions from the 
gravitational Chern-Simons term vanish and we must arrive
at the ``self dual" solutions of the Einstein-Maxwell-Chern-Simons
theory which was studied earlier in \cite{DerObu}. Taking
\( k \m \rightarrow \infty\) in (\ref{Omega}), one gets
\be
\O \;\; \rightarrow \;\; c_0 + \frac{v_0 l}{2} e^{-2 \r /l} 
- \frac{1}{2 k m y_0 (k m -1/l)} e^{2(k m -1/l)\r} \;\;\; .
\ee
To compare this result with \cite{DerObu}, take equation (23) of
\cite{DerObu} and substitute in (21) to find
\be
\O = c_0 + \frac{l}{2} e^{-2 \r /l} 
+ \frac{u_0}{2 (k m -1/l)} e^{2(k m -1/l)\r} \;\;\; . 
\ee 
Hence our solution has the correct limit provided
\( v_0 =1 \) and \( u_0 = - \frac{1}{k m y_0} \).

As for the physical properties of our solution, it is obvious that
depending on the values of the integration constants $c_0$, $v_0$ and
$y_0$, and also on the values of $l$, $\m$ and $m$, the metric functions
may have singularities. One needs to carefully study the causal structure
of our solution to understand its nature and the geometry that it describes. 
However, just as explained in \cite{DS}, this is again rendered impossible 
since one cannot invert the functional relation \( r = h(\r) \) which is
crucial to put the line element given by (\ref{metes}), (\ref{met}) and
(\ref{solf})..(\ref{sola}) into the well studied form of the BTZ (and hence
the AdS) metric.

Nevertheless one can still work out the quasilocal mass and the angular 
momentum corresponding to this solution. The analysis is very similar
to the ones given in \cite{DerObu} and \cite{DS}, and we refer the reader
to these articles for the missing details below. The quasilocal angular
momentum is
\be
j(r)=k h_0^2 \vp(r) \;\; ,
\ee
whereas the quasilocal mass is given by
\be
m(r)= a_0 j(r) = k a_0 h_0^2 \vp(r) \;\; 
\ee
in an AdS background. Here $\vp$ is as given in (\ref{varphi}) and it is
understood that \( r = h(\r) \) has to be inverted so that $\vp$ is a 
function of $r$.

The total angular momentum $J$ and the total mass $M$ are defined by
the limits \( J \equiv j(r) |_{r \rightarrow \infty} \) and
\( M \equiv m(r) |_{r \rightarrow \infty} \), respectively. We assume
that $r \rightarrow \infty$ limits can be found by taking the
$\r \rightarrow \infty$ limits in our solution, i.e. that
\( \vp(r)|_{r \rightarrow \infty} = \vp(\r)|_{\r \rightarrow \infty} \).
Just as was done in \cite{DerObu} and \cite{DS}, start by examining $a(r)$.
Depending on the values of $l$, $\m$ and $m$, $a$ either goes to $-a_0$ or
$-a_0+ \frac{k f_0}{h_0 c_0}$ as $r \rightarrow \infty$. For
$a$ to vanish asymptotically as $r \rightarrow \infty$, $a_0$ has to 
be chosen either as 0 or as $\frac{k f_0}{h_0 c_0}$. When $1/l > k \m$ or
$k m > 0$, $a_0=0$ and hence $M=0$ whereas $J \rightarrow \infty$. 
For $1/l < k \m$ and $k m < 0$, $J= k h_0^2 v_0$. Then 
$a_0 = \frac{k f_0}{h_0 c_0}$, $M=a_0 J$ and both $M$ and $J$ are finite. 
As $k \m \rightarrow \infty$, this solution approaches the extremal BTZ
solution. We again refer the reader to \cite{DerObu} and \cite{DS} 
for the details.

We next examine the special cases i) $m=0$ and $1/l \neq 0$,
ii) $m \neq 0$ and $1/l = 0$, and iii) $m= 1/l = 0$. In all these
cases, one has to go back to the original equations since simply
taking the corresponding $m \rightarrow 0$ and (or) $1/l \rightarrow 0$
limits in the above expressions do not give the correct answers.

i) $m=0$ and $1/l \neq 0$:

In this case, the equation for $v$ becomes
\be
v^\pr + \left(  
\frac{ y_0 (1/l - \m k) e^{(1/l - \m k ) \r} }{1 + y_0 e^{(1/l -\m k ) \r} }
\right) v = 2 \;\;\;  \label{veqnm=0}
\ee
which yields
\be
v = \frac{1}{1+ y_0 e^{(1/l - \m k) \r} } \;
\left( v_0 + \frac{2}{y_0} \r + \frac{2}{1/l- \m k} e^{(1/l-\m k)\r} 
\right) \;\;\; \label{vsolnm=0}
\ee
with an integration constant $v_0$.

Again using the definition of $u$, (\ref{defnu}), one gets a new differential
equation for $\b$ and by setting \( \o \equiv \frac{1}{\b} \), one again
finds a differential equation of the form (\ref{omeqn}). Only this time
\be 
\vp = v_0 + \frac{2}{y_0} \r + \frac{2}{1/l- \m k} e^{(1/l-\m k) \r} 
 \;\;\; , \label{varphim=0}
\ee
and 
\be
\O = c_0 + \frac{v_0 l}{2} e^{-2 \r /l} + \frac{2}{1/l^2 - \m^2} 
e^{-(1/l + \m k) \r} + \frac{l}{y_0} \; \r \; e^{-2 \r/l}
\;\;\; . \label{Omegam=0}
\ee

The metric functions are again of the form (\ref{solf})..(\ref{sola}) only
that $\O$ is now given by (\ref{Omegam=0}). Now the magnetic field becomes
\be
B^2 = \frac{\m k - 1/l}{k \m y_0} \; e^{-2 \r / l} 
\; \O^{-1} \; .
\ee

ii) $m \neq 0$ and $1/l = 0$:

For this case, following similar steps as was done in i), one gets   
\be
\vp = v_0 - \frac{2}{k \m} e^{-k \m \r} + \frac{1}{k m y_0} 
e^{2 k m \r} \;\;\; , 
\ee
and
\be
\O = c_0 - v_0 \r - \frac{2}{\m^2} e^{- k \m \r} - 
\frac{1}{2 y_0 m^2} e^{2 k m \r} \;\;\; .
\ee

Using these, the metric functions are then found to be
\begin{eqnarray}
f & = & f_0 \; \O^{-1/2} \;\;\; , \label{solfl=0} \\
h & = & h_0 \; \O^{1/2} \;\;\; , \label{solhl=0} \\
a & = & -a_0 + k \; \frac{f_0}{h_0} \; \O^{-1} \;\;\; , \label{solal=0}
\end{eqnarray}
for some new integration constants $f_0$, $h_0$, $a_0$ whereas the
magnetic field becomes
\be
B^2 = \frac{\m + 2 m}{\m y_0} \; e^{2k m \r} 
\; \O^{-1} \; .
\ee

iii) $m= 1/l = 0$:

In this case, the equation for $v$ is
\be
v^\pr - \left(  
\frac{k \m y_0 e^{-k \m \r} }{1 + y_0 e^{-k \m \r} }
\right) v = 2 \;\;\;  \label{veqnlm=0}
\ee
which is easily integrated as
\be
v = \frac{1}{1+ y_0 e^{-k \m \r} } \;
\left( v_0 + \frac{2}{y_0} \r - \frac{2}{k \m} e^{-k \m \r} 
\right) \;\;\; . \; \label{vsolnlm=0}
\ee
Following similar steps as in i) and ii), one finds
\be 
\vp = v_0 + \frac{2}{y_0} \r - \frac{2}{k \m} e^{-k \m \r} 
 \;\;\; , \label{varphilm=0}
\ee
and 
\be
\O = c_0 - v_0 \r - \frac{1}{y_0} \r^2 - \frac{2}{\m^2} e^{-k \m \r}
\;\;\; . \label{Omegalm=0}
\ee
Now the metric functions are of the form (\ref{solfl=0})..(\ref{solal=0})
where it is understood that $\O$ is given by (\ref{Omegalm=0}). Finally
the magnetic field is simply
\be
B^2 = \frac{1}{y_0} \O^{-1} \;\;\; .
\ee

\bigskip
In this work, we have obtained a general class of ``self dual" solutions to
TMG model coupled with Maxwell-Chern-Simons theory which covers all the 
particular cases studied previously. Even though we couldn't give a detailed
analysis of the causal structure of our solutions, we were able to analyze
the corresponding total angular momentum and the total mass. We also found
the special solutions corresponding to taking the cosmological constant and
(or) the Chern-Simons coupling constant to zero.

\newpage
%\begin{thebibliography}{99}

%\end{thebibliography}
\end{document}